\documentclass{article}
\usepackage{authblk}
\usepackage{amsmath}
\usepackage{graphicx}
\usepackage{hyperref} 
\usepackage[numbers,sort&compress]{natbib}
\usepackage{quantikz}
\usepackage{subcaption}
\usepackage{adjustbox}
\usepackage{tikz}
\usepackage{makecell}

\newcommand{\equalcontribsymbol}{\textsuperscript{\ensuremath{\dagger}}}
\newcommand{\equalcontribfootnote}{%
  \begingroup
  \renewcommand\thefootnote{}
  \footnotetext{%
    \equalcontribsymbol~These authors contributed equally to this work.%
  }
  \addtocounter{footnote}{-1}
  \endgroup
}

\title{Noise-Resilient Quantum Chemistry with Half the Qubits}

\author[1]{Shane McFarthing\thanks{Corresponding author: \href{mailto:mcfarthing.s@qunovacomputing.com}{mcfarthing.s@qunovacomputing.com}}\equalcontribsymbol}

\author[1]{Aidan Pellow-Jarman\equalcontribsymbol}
\author[1, 2, 3]{Francesco Petruccione}

\affil[1]{Qunova Computing Inc.}
\affil[2]{School of Data Science and Computational Thinking, Stellenbosch University, Stellenbosch, South Africa}
\affil[3]{National Institute for Theoretical and Computational Sciences (NITheCS), South Africa}

\date{}

\begin{document}

\maketitle
\equalcontribfootnote

\begin{abstract}
Sample-based quantum diagonalization (SQD) offers a powerful route to accurate quantum chemistry on noisy intermediate-scale quantum (NISQ) devices by combining quantum sampling with classical diagonalization. Here we introduce HSQD, a novel half-qubit SQD approach that halves the qubit requirement for simulating a chemical system and drastically reduces overall circuit depth and gate counts, suppressing hardware noise. When modeling the dissociation of the nitrogen molecule with a (10e, 26o) active space, HSQD matches the accuracy of SQD on IBM quantum hardware using only half the number of qubits and 40\% fewer measurements. We further enhance HSQD with a heat-bath configuration interaction (HCI) inspired selection of the samples, forming HCI-HSQD. This yields sub-millihartree accuracy across the N$_2$ potential energy surface and produces subspaces up to 39\% smaller than those from classical HCI, showing a significant improvement in the compactness of the ground-state representation. Finally, we demonstrate HCI-HSQD's scalability using iron–sulfur clusters, reaching active spaces of up to (54e, 36o) while using only half as many qubits as the original SQD. For these systems, HCI-HSQD reduces SQD energy errors by up to 76\% for [2Fe–2S] and 26\% for [4Fe–4S], while also reducing subspace sizes, halving measurement requirements, and eliminating expensive post-processing.
Together, these results establish half-qubit SQD as a noise-resilient and resource-efficient pathway toward practical quantum advantage in strongly correlated chemistry.

\end{abstract}

\section{Introduction}

Performing accurate quantum chemistry simulations is a central challenge in materials science and drug discovery, where predicting molecular properties can drive innovation. A key example is the calculation of molecular ground states, which determine stability and reactivity. Unfortunately, the computational cost of simulating these systems scales exponentially, rendering classical methods infeasible for most chemically relevant molecules. Quantum computers, by contrast, naturally represent quantum states and thus offer the prospect of exponential efficiency gains in simulating quantum systems. Recent algorithmic advances have aimed to harness near-term quantum devices for this purpose.

Among these, quantum selected configuration interaction (QSCI) has emerged as a promising hybrid quantum–classical framework \cite{Kanno2023, Nakagawa2024, Robledo2025, PellowJarman2025,Sugisaki2025, Mikkelsen2025, Yu2025, Shirakawa2025, Yoo2026}. Unlike variational quantum eigensolver (VQE) \cite{Peruzzo2014} approaches, QSCI methods avoid expensive measurement overheads while maintaining accuracy on noisy intermediate-scale quantum (NISQ) hardware. Several variants have been proposed: some generate a single quantum state for sampling \cite{Kanno2023, Robledo2025}, while others iteratively update states to refine the sampled subspace \cite{PellowJarman2025, Shirakawa2025, Yoo2026}. In all cases, the goal is to construct a compact basis of electron configurations that contains the ground state support. Recent demonstrations show that QSCI methods can simulate molecules of up to 72 spin-orbitals on current quantum hardware with significantly improved accuracy compared to traditional NISQ strategies \cite{Robledo2025, Shirakawa2025, Yoo2026}. Despite this progress, a fundamental obstacle remains in generating high-quality electron configurations using quantum hardware.

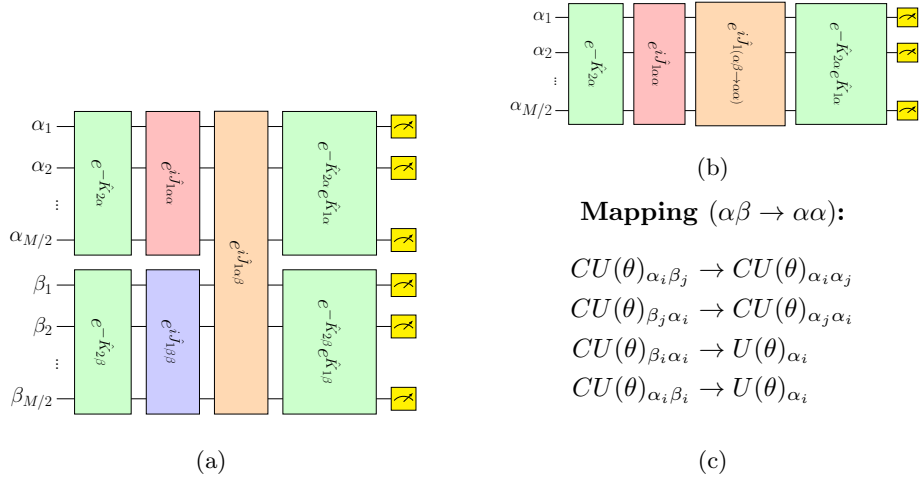
\begin{figure*}[t]
    \centering

    \begin{subfigure}[b]{0.45\textwidth}
        \centering
        \begin{adjustbox}{max width=\textwidth}
            \begin{quantikz}[font=\huge]
                \lstick{$\alpha_1$} & \gate[4, style={fill=green!20}, label style={rotate=-90}]{\text{$e^{-\hat{K}_{2\alpha}}$}} & \gate[4,  style={fill=red!25}, label style={rotate=-90}]{\text{$e^{i\hat{J}_{1\alpha\alpha}}$}} & \gate[8, style={fill=orange!30}, label style={rotate=-90}]{\text{$e^{i\hat{J}_{1\alpha\beta}}$}} & \gate[4, style={fill=green!20}, label style={rotate=-90}]{\text{$e^{-\hat{K}_{2\alpha}}e^{\hat{K}_{1\alpha}}$}} & \meter[style={fill=yellow}]{} \\
                \lstick{$\alpha_2$} & & & & & \meter[style={fill=yellow}]{} \\
                \vdots \\
                \lstick{$\alpha_{M/2}$}  & & & & & \meter[style={fill=yellow}]{} \\
                \lstick{$\beta_1$}  &\gate[4, style={fill=green!20}, label style={rotate=-90}]{\text{$e^{-\hat{K}_{2\beta}}$}} & \gate[4, style={fill=blue!20}, label style={rotate=-90}]{\text{$e^{i\hat{J}_{1\beta\beta}}$}} & & \gate[4, style={fill=green!20}, label style={rotate=-90}]{\text{$e^{-\hat{K}_{2\beta}}e^{\hat{K}_{1\beta}}$}} & \meter[style={fill=yellow}]{} \\
                \lstick{$\beta_2$}  & & & & & \meter[style={fill=yellow}]{} \\
                \vdots \\
                \lstick{$\beta_{M/2}$}  & & & & & \meter[style={fill=yellow}]{} \\
            \end{quantikz}
        \end{adjustbox}
        \caption{}
        \label{fig:1a}
    \end{subfigure}
    \hfill
    \begin{subfigure}[b]{0.45\textwidth}
        \centering
        \begin{subfigure}[b]{\textwidth}
            \centering
            \begin{adjustbox}{max width=\textwidth}
                \begin{quantikz}[font=\Huge]
                \lstick{$\alpha_1$} & \gate[4, style={fill=green!20},label style={rotate=-90}]{\text{$e^{-\hat{K}_{2\alpha}}$}} & \gate[4, style={fill=red!25}, label style={rotate=-90}]{\text{$e^{i\hat{J}_{1\alpha\alpha}}$}} & \gate[4, style={fill=orange!30}, label style={rotate=-90}]{\text{$e^{i\hat{J}_{1(\alpha\beta\rightarrow\alpha\alpha)}}$}} & \gate[4, style={fill=green!20}, label style={rotate=-90}]{\text{$e^{-\hat{K}_{2\alpha}}e^{\hat{K}_{1\alpha}}$}} & \meter[style={fill=yellow}]{} \\
                \lstick{$\alpha_2$} & & & & & \meter[style={fill=yellow}]{} \\
                \vdots \\
                \lstick{$\alpha_{M/2}$}  & & & & & \meter[style={fill=yellow}]{} \\
            \end{quantikz}
            \end{adjustbox}
            \caption{}
            \label{fig:1b}
        \end{subfigure}

        \vspace{0.5em} 

        \begin{subfigure}[b]{\textwidth}
            \centering
    		\textbf{Mapping $\left(\alpha\beta \rightarrow \alpha\alpha\right)$:}
    		\[
    			\begin{aligned}
    				CU(\theta)_{\alpha_i\beta_j} &\rightarrow CU(\theta)_{\alpha_i\alpha_j} \\
    				CU(\theta)_{\beta_j\alpha_i} &\rightarrow CU(\theta)_{\alpha_j\alpha_i} \\
				CU(\theta)_{\beta_i\alpha_i} &\rightarrow U(\theta)_{\alpha_i} \\
				CU(\theta)_{\alpha_i\beta_i} &\rightarrow U(\theta)_{\alpha_i}
    			\end{aligned}
   		 \]            
    	    \caption{}
            \label{fig:1c}
        \end{subfigure}
    \end{subfigure}
    \caption{\textbf{Half-qubit LUCJ circuit.} \textbf{(a)} Original $M$-qubit truncated LUCJ circuit used in SQD \cite{Robledo2025}, comprising orbital rotation unitaries $\exp(-\hat{K}_{\mu\sigma})$, same-spin cluster operators $\exp(i(\hat{J}_{\mu\alpha\alpha} + \hat{J}_{\mu\beta\beta}))$, and opposite-spin cluster operators $\exp(i\hat{J}_{\mu\alpha\beta})$, where $\mu$ is the layer index, $\sigma \in \{\alpha,\beta\}$ denotes spin, $\alpha$ refers to spin-up orbitals, and $\beta$ to spin-down orbitals. \textbf{(b)} Modified $M/2$-qubit LUCJ circuit, in which opposite-spin cluster operators $\exp(i\hat{J}_{\mu\alpha\beta})$ are remapped to same-spin operators $\exp(i\hat{J}_{\mu\alpha\alpha})$, represented as $\exp(i\hat{J}_{\mu(\alpha\beta\rightarrow\alpha\alpha)})$, with $\alpha_i$ and $\beta_i$ denoting the spin-up and spin-down orbitals of spatial orbital $i$. \textbf{(c)} The circuit remapping used to obtain $\exp(i\hat{J}_{\mu(\alpha\beta\rightarrow\alpha\alpha)})$, showing parameterized single-qubit gates $U(\theta)_{\sigma_i}$ on qubit $\sigma_i$ and controlled gates $CU(\theta)_{\sigma_i\gamma_j}$ with control qubit $\sigma_i$ and target qubit $\gamma_j$, where $\sigma, \gamma \in \{\alpha,\beta\}$ and $\sigma_i, \gamma_j$ are the qubits of the corresponding spin-orbitals for spatial orbitals $i$ and $j$, respectively.}
    \label{fig:1}
\end{figure*}

\begin{figure*}[ht]
    \centering
    \includegraphics[width=0.8\textwidth]{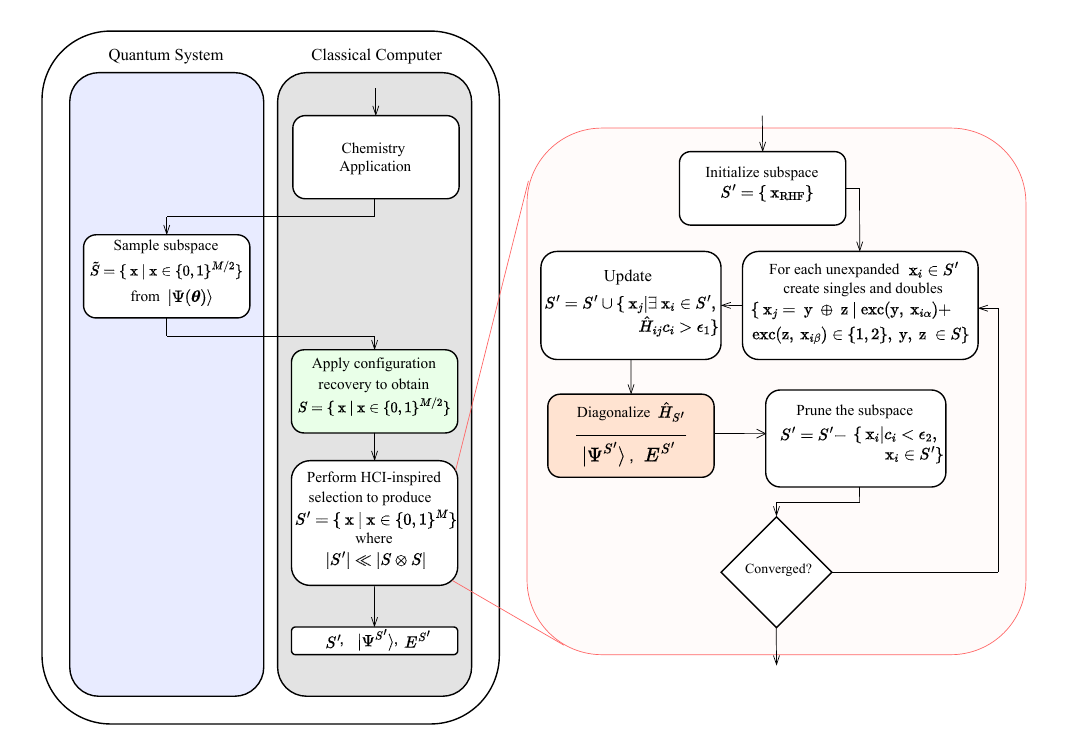}
    \caption{\textbf{Workflow of the HCI-HSQD method.} \textbf{Left:} A noisy subspace $\tilde{S}$ is sampled from a variational state $\lvert\Psi(\boldsymbol\theta)\rangle$ prepared on a quantum processor via a $M/2$-qubit circuit, reducing the qubit requirement by half. A classical recovery step produces the corrected subspace $S$, which is passed to an HCI-inspired selection to yield a compact subspace $S'$ and its ground state $\lvert\Psi^{S'}\rangle$ with energy $E^{S'}$. \textbf{Right:} Construction of $S'$ from $S$ without explicit formation of the full tensor-product space $S_{\text{tensor}}=S\otimes S$, by dynamically combining half-configurations and filtering them with the standard HCI criterion \cite{Holmes2016}.}
    \label{fig:2}
\end{figure*}

\begin{figure*}[th]
    \centering
    \begin{subfigure}[b]{0.46\textwidth}

        \centering
        \includegraphics[width=\textwidth]{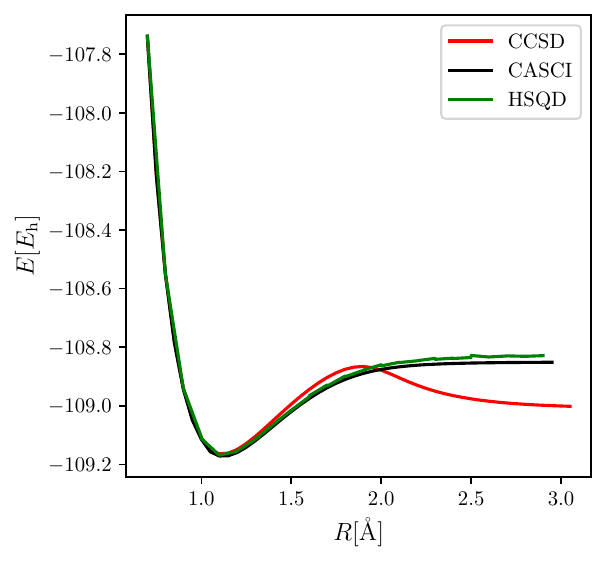}
        \caption{}
        \label{fig:sub1}
    \end{subfigure}
    \begin{subfigure}[b]{0.46\textwidth}
        \centering
        \includegraphics[width=\textwidth]{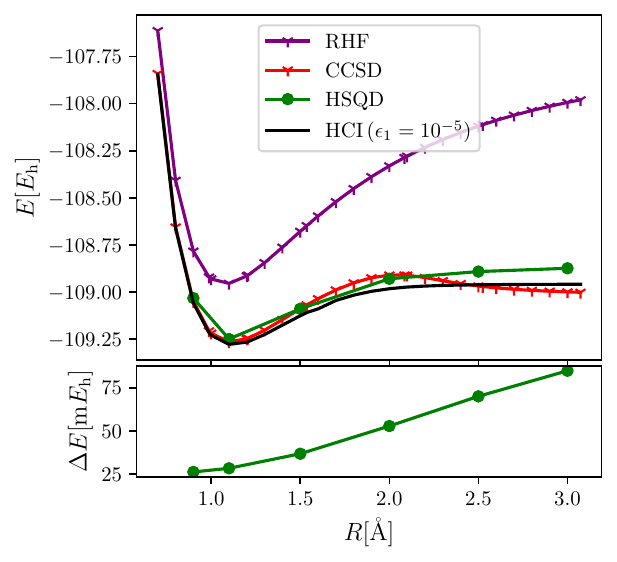}
        \caption{}
        \label{fig:sub2}
    \end{subfigure}

    \caption{\textbf{Chemical systems can be modeled accurately with half-qubit circuits.} \textbf{(a)} Energies from the modified half-qubit LUCJ ansatz on a noiseless simulator for N$_2$ in the $6$-$31$G basis with a (10e, 12o) active space. The remapped LUCJ circuit yields lower energies than CCSD, indicating that the remapped cluster operators $\exp(i\hat{J}{\mu(\alpha\beta\rightarrow\alpha\alpha)})$ can accurately capture correlations within the HSQD framework. \textbf{(b)} Potential energy surface of N$_2$ in the cc-pVDZ basis with a (10e, 26o) active space, comparing HSQD to classical HCI. HSQD reproduces the dissociation curve with accuracy comparable to the original SQD \cite{Robledo2025} while using half the qubits and 40\% fewer quantum measurements.}
    \label{fig:3}
\end{figure*}

At the heart of both classical and quantum SCI methods lies the construction of a suitable subspace of electron configurations. For QSCI methods, this requires the quantum processor to efficiently generate the most important configurations. Two key factors determine success: the quality of the quantum state prepared on the device, and the resilience of this state to noise. If the wave function of the prepared state does not contain the electron configurations in the ground-state support, then sampling will inevitably miss them, regardless of device quality. Careful ansatz and parameter choices are therefore crucial, with strategies ranging from directly approximating the ground state \cite{Kanno2023, Nakagawa2024, Robledo2025, Shirakawa2025} to constructing auxiliary states that enable efficient sampling \cite{PellowJarman2025, Sugisaki2025, Mikkelsen2025, Yu2025, Yoo2026}.

Noise presents a second, equally significant challenge. Even if the exact ground-state wave function were prepared, device noise can corrupt key configurations and break system symmetries \cite{Robledo2025, Reinholdt2025}. Furthermore, hardware limitations restrict the depth of viable ansätze, excluding many chemically inspired ansätze, such as the unitary coupled-cluster with singles and doubles (UCCSD) \cite{Bartlett1989}, and k-UpCCGSD \cite{Lee2019}, that are too deep for NISQ execution. This makes balancing expressivity and feasibility particularly difficult. To address these challenges, several corrective and algorithmic strategies have been proposed.

One prominent approach is the use of self-consistent configuration recovery (SCCR) \cite{Robledo2025} to correct invalid samples obtained from symmetry-preserving ansätze. SCCR adjusts the spin occupations of erroneous configurations towards a reference occupation, iteratively refining the sample distribution to more closely approximate the ground-state. This strategy has proven effective in recent studies where it helped recover useful quantum signals on noisy devices \cite{Robledo2025, Shirakawa2025}.

Nonetheless, SCCR alone does not resolve the scarcity of valid samples. Increasing measurement counts can mitigate this issue, but the cost grows rapidly, particularly for strongly correlated systems where large numbers of low-amplitude configurations cumulatively play an important role \cite{Reinholdt2025}. Complementary progress has focused on improving configuration interaction (CI) solvers \cite{Holmes2016, Smith2017, Sharma2017, Xu2025}. Modern solvers often exploit tensor-product steps that enlarge the sampled subspace, both enhancing accuracy and enabling computational shortcuts that dramatically accelerate diagonalizations. Building on these developments, we propose a variant of sampled-based quantum diagonalization (SQD) \cite{Robledo2025} that further reduces noise sensitivity and hardware requirements.

In this work, we introduce the half-qubit SQD (HSQD) method that halves the number of qubits needed to simulate chemical systems with the Jordan–Wigner mapping. Our method employs a modified local unitary cluster Jastrow (LUCJ) ansatz \cite{Motta2023} that simulates same-spin correlations and recovers inter-spin correlation through a tensor-product step in the CI solver. This approach halves the number of qubits and reduces circuit depths by roughly 50\%. This limits noise accumulation during state preparation and quantum sampling, and increases the size of the chemical systems that can be simulated on a given quantum computer.

\begin{figure*}[th]
    \centering
    \begin{subfigure}[b]{0.46\textwidth}

        \centering
        \includegraphics[width=\textwidth]{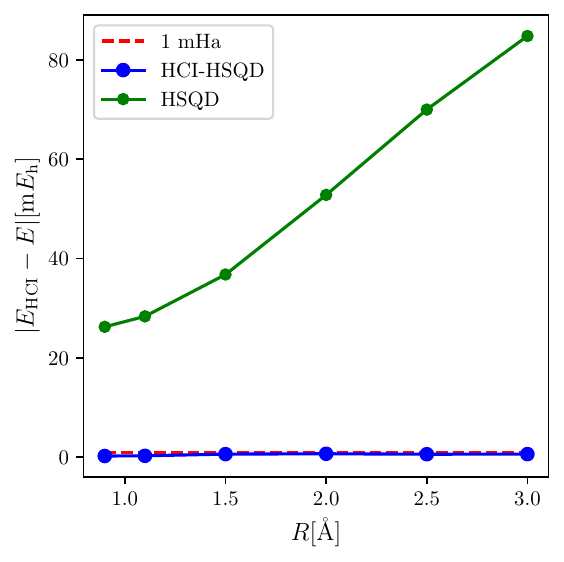}
        \caption{}
        \label{fig:sub1}
    \end{subfigure}
    \begin{subfigure}[b]{0.46\textwidth}
        \centering
        \includegraphics[width=\textwidth]{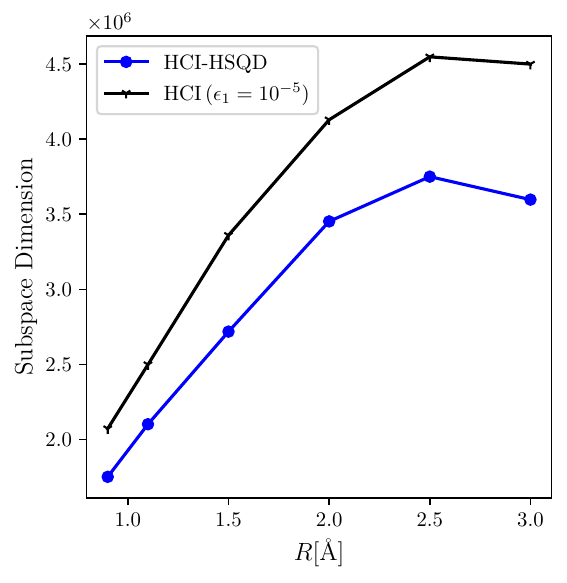}
        \caption{}
        \label{fig:sub2}
    \end{subfigure}

    \caption{\textbf{Improved selection of sampled configurations with HCI-HSQD yields chemical accuracy.} \textbf{(a)} Energy errors across the N$_2$ (cc-pVDZ, 10e, 26o) potential energy surface. Replacing the stochastic subspace selection \cite{Robledo2025} in HSQD with the HCI-inspired selection developed in this work reduces the energy error of half-qubit samples to below 1 mHa across all geometries. \textbf{(b)} Comparison of subspace sizes returned by classical HCI and HCI-HSQD. HCI-HSQD produces subspaces that are up to 39\% more compact, demonstrating that half-qubit sampled configurations accurately capture the ground-state support and can accelerate selected CI expansions.}
    \label{fig:4}
\end{figure*}

\begin{figure*}[th]
    \centering
    \begin{subfigure}[b]{0.46\textwidth}

        \centering
        \includegraphics[width=\textwidth]{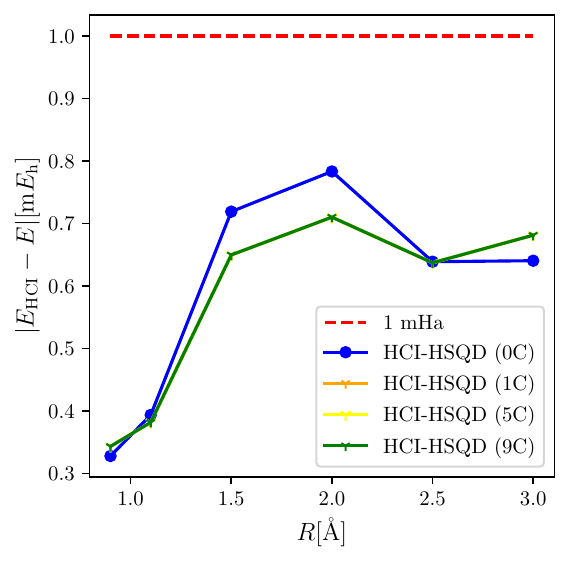}
        \caption{}
        \label{fig:sub1}
    \end{subfigure}
    \begin{subfigure}[b]{0.46\textwidth}
        \centering
        \includegraphics[width=\textwidth]{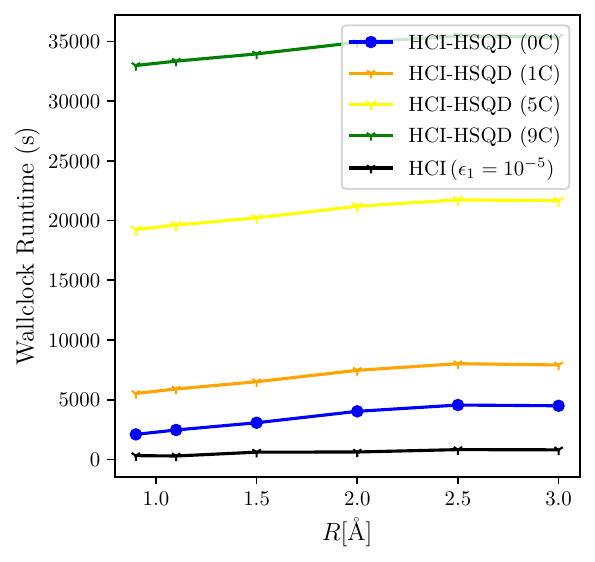}
        \caption{}
        \label{fig:sub2}
    \end{subfigure}

    \caption{\textbf{Cost to correct noisy samples can be reduced with HCI-HSQD} \textbf{(a)} Energies along the N$_2$ (cc-pVDZ, 10e, 26o) potential energy surface using HCI-HSQD with different configuration recovery procedures applied to the noisy samples $\tilde{S}$. Labels (0C), (1C), (5C), and (9C) indicate corrections using the occupation vector \textbf{n} from a single diagonalization of valid configurations in $\tilde{S}$, 1 cycle of SCCR, 5 cycles of SCCR, and 9 cycles of SCCR, respectively. SCCR slightly lowers the energies, but all corrections yield errors below 1 mHa relative to classical HCI. \textbf{(b)} Wallclock runtimes with each correction method in (a), with classical HCI included as a reference. SCCR significantly increases runtime, so the modest energy improvement must be weighed against computational cost.}
    \label{fig:5}
\end{figure*}

By reducing the number of qubits, the circuit depth, and the number of gates, our half-qubit framework extends the applicability of QSCI methods to larger basis sets and more strongly correlated systems under NISQ constraints. It also improves the fidelity of quantum-generated samples and reduces the need for expensive post-processing. This lays the foundation for scalable, noise-resilient quantum–classical simulations.

We first validate our approach by reproducing the modeling of the dissociation of the nitrogen molecule in the cc-pVDZ basis with a (10 electrons, 26 orbitals) active space. Our method achieves a similar accuracy across the potential energy surface (PES) compared to SQD \cite{Robledo2025}, but requires at least 40\% fewer quantum measurements and half the number of qubits. These results show that reducing qubit counts and circuit depth can yield significant accuracy and efficiency improvements for QSCI methods on NISQ hardware.

\begin{figure*}[th]
    \centering
    \begin{subfigure}[b]{0.46\textwidth}

        \centering
        \includegraphics[width=\textwidth]{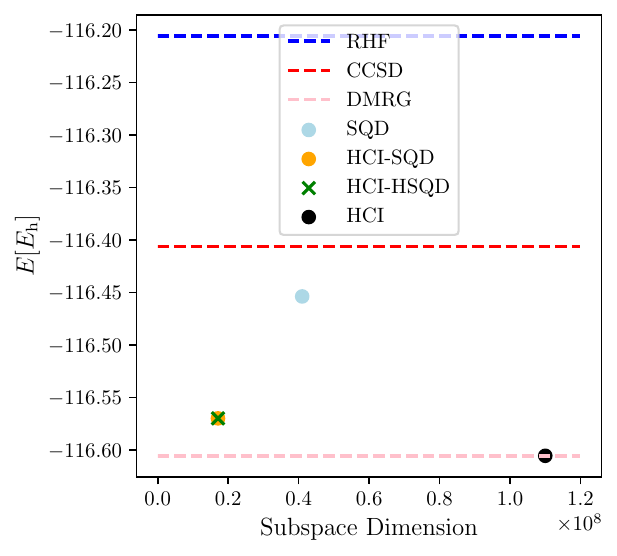}
        \caption{}
        \label{fig:fes2}
    \end{subfigure}
    \begin{subfigure}[b]{0.46\textwidth}
        \centering
        \includegraphics[width=\textwidth]{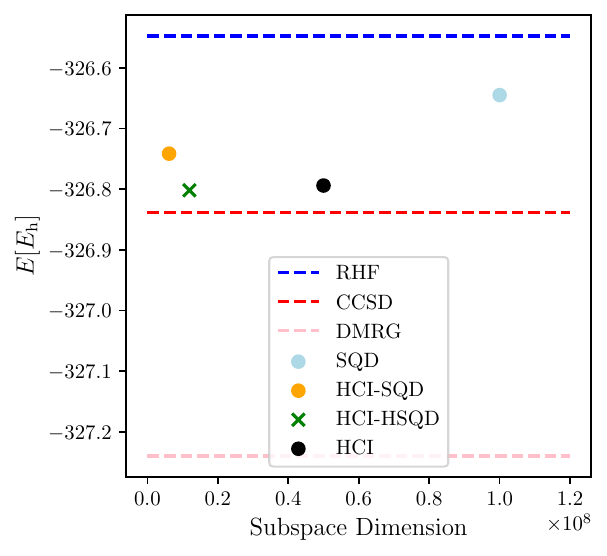}
        \caption{}
        \label{fig:fes4}
    \end{subfigure}

    \caption{\textbf{Improved accuracy and efficiency on large, correlated systems.} \textbf{(a)} $[2\text{Fe}$–$2\text{S}]$ cluster with a (30e, 20o) active space, and \textbf{(b)} $[4\text{Fe}$–$4\text{S}]$ cluster with a (54e, 36o) active space. Plotted are energies (Hartree) as a function of subspace dimension for RHF, CCSD, HCI, SQD, HCI-SQD (SQD with HCI-inspired selection), and HCI-HSQD (half-qubit SQD with HCI-inspired selection). A simple probability-based correction is applied to the noisy quantum samples $\tilde{S}$ in HCI-SQD and HCI-HSQD. For $[2\text{Fe}$–$2\text{S}]$, HCI-HSQD and HCI-SQD give identical energies, reducing the SQD energy error by 76\% with a 58\% smaller subspace. However, HCI-HSQD used only half the qubits and measurements as SQD and HCI-SQD. For $[4\text{Fe}$–$4\text{S}]$, HCI-HSQD outperforms both SQD and HCI-SQD, reducing the SQD error by 26\% with an 88\% smaller subspace. HCI-HSQD also outperforms classical HCI, reducing its energy error by ~1.7\% with a 76\% smaller subspace.}
    \label{fig:fes}
\end{figure*}

We further enhance HSQD by refining the stochastic subspace construction proposed in SQD \cite{Robledo2025}. The standard Monte Carlo estimator provides a useful approximation of ground-state orbital occupations in large subspaces \cite{Robledo2025}, but its stochastic nature does not guarantee an optimal subspace selection. To address this, we introduce an HCI-inspired deterministic selection that identifies the most significant determinants from within the corrected half-qubit quantum samples. We call this combined strategy HCI-HSQD and applying it to the nitrogen system yields energy errors below $1$ mHa relative to HCI across the PES and produces subspaces that are 18–39\% more compact than HCI. This result highlights the efficiency of our method and the value of generating subspaces of electron configurations through the sampling of quantum states. We also investigate our method's dependency on the type of configuration recovery applied to the noisy samples and find that for the same accuracy, we can greatly reduce the cost of the correction procedure. 

Finally, we use HCI-HSQD to reproduce large-scale SQD simulations \cite{Robledo2025} of the synthetic iron–sulfur clusters $[\text{Fe}_2\text{S}_2(\text{SCH}_3)_4]^{2-}$ and $[\text{Fe}_4\text{S}_4(\text{SCH}_3)_4]^{2-}$ \cite{Sharma2014}, hereafter denoted $[2\text{Fe}$–$2\text{S}]$ and $[4\text{Fe}$–$4\text{S}]$. For $[2\text{Fe}$–$2\text{S}]$ (30 electrons, 20 orbitals), HCI-HSQD reduces the SQD energy error relative to the classical density matrix renormalization group (DMRG) \cite{White1992, White1993} reference by 76\%, using a 58\% smaller subspace. For $[4\text{Fe}$–$4\text{S}]$ (54 electrons, 36 orbitals), the SQD error is reduced by 26\% with an 88\% smaller subspace. For both systems, HCI-HSQD requires half as many qubits and measurements as SQD and eliminates the need for SCCR, replacing it with a single non-iterative correction that does not require any diagonalizations. This represents a substantial increase in the accuracy and efficiency of the simulations, greatly reducing the amount of computational resources required for such applications.

\section{Results}

QSCI methods \cite{Kanno2023, Nakagawa2024, Robledo2025, PellowJarman2025,Sugisaki2025, Mikkelsen2025, Yu2025, Shirakawa2025, Yoo2026} approximate the ground state of chemical systems by solving the time-independent Schrödinger equation under the Born–Oppenheimer approximation,

\begin{equation}
    \hat{H}\vert\Psi\rangle = E\vert\Psi\rangle,
\end{equation}
where the electronic Hamiltonian $\hat{H}$ is expressed as

\begin{equation}
    \hat{H} = \sum_{pq}h_{pq}\hat{a}^\dagger_p\hat{a}_q + \tfrac{1}{2}\sum_{pqrs}h_{pqrs}\hat{a}^\dagger_p\hat{a}^\dagger_q\hat{a}_r\hat{a}_s.
\end{equation}
Here, \(h_{pq}\) and \(h_{pqrs}\) are the one- and two-electron integrals over spin-orbitals \(p, q, r, s\), and $\hat{a}^\dagger$ and $\hat{a}$ are the fermionic creation and annihilation operators, respectively.

By repeatedly measuring the quantum state \(\vert\Psi\rangle\) in the computational basis, a subspace of electron configurations \(S = \{\text{x } \vert \text{ x}\in \{0, 1\}^M\}\) is obtained, where \(M\) is the number of spin-orbitals. The subspace Hamiltonian \(\hat{H}_S\) is constructed by projecting the full Hamiltonian $\hat{H}$ onto the subspace spanned by \(S\),

\begin{equation}
    \hat{H}_S = \hat{P}_S\hat{H}\hat{P}_S, \quad \hat{P}_S = \sum_{\text{x}\in S}\vert\text{x}\rangle\langle\text{x}\vert,
\end{equation}
where $\vert\text{x}\rangle$ is the computational basis state represented by the binary string x~$ \in S$.

The approximate ground state \(\vert\Psi^S\rangle\) and energy \(E^S\) are then obtained through the exact diagonalization of \(\hat{H}_S\), which is feasible due to the reduced dimension of \(S\) compared to the original Hilbert space. Provided that the sampled subspace \(S\) contains the electron configurations in the true ground-state support, this approach yields accurate approximations of the exact ground-state and its energy.

Noise on prefault-tolerant devices can distort the prepared quantum state \(\vert\Psi\rangle\), producing a noisy subspace \(\tilde{S}\) containing electron configurations that may violate physical symmetries, such as the total particle number. To address this, SCCR \cite{Robledo2025} was developed to iteratively correct invalid configurations in \(\tilde{S}\) using a reference orbital occupation vector \(\mathbf{n}\) that is refined in each cycle. In every SCCR cycle, \(k\) batches of electron configurations \(S^{(k)}\) are subsampled from the corrected subspace \(S\). The corresponding subspace Hamiltonians \(\hat{H}_{S^{(k)}}\) are diagonalized to produce ground states \(\vert\Psi^{(k)}\rangle\), energies \(E^{(k)}\), and orbital occupations \(\mathbf{n}^{(k)}\). The reference occupation \(\mathbf{n}\) is then updated as the mean of the computed occupancies \(\mathbf{n}^{(k)}\), and the next cycle begins with the new corrected subspace.

The SQD framework has been applied to systems of up to 72 spin-orbitals and 54 electrons, demonstrating that QSCI methods can produce ground-state approximations of higher accuracy than non-QSCI NISQ approaches within a quantum-centric supercomputing workflow \cite{Robledo2025}. The quantum state \(\vert\Psi\rangle\) in SQD \cite{Robledo2025} was prepared using a LUCJ ansatz \cite{Motta2023},

\begin{equation}
    \vert\Psi\rangle = \sum_{\mu=1}^Le^{\hat{K}_{\mu}}e^{i\hat{J}_\mu}e^{-\hat{K}_\mu}\vert\Psi_\text{RHF}\rangle,
\end{equation}
where $\vert\Psi_\text{RHF}\rangle$ is the Hartree-Fock reference state, $\hat{K}_\mu=\sum_{pr,\sigma}K_{pr}^\mu\hat{a}^\dagger_{p\sigma}\hat{a}_\sigma$ are one-body operators, and $\hat{J}_\mu = \sum_{pr,\sigma\tau}J_{p\sigma,r\tau}^\mu\hat{n}_{p\sigma}\hat{n}_{r\tau}$ are density-density operators mapped onto adjacent qubits.

Due to device constraints, a single layer ($L=1$) was used and an additional one-body operator was included to mitigate over-concentration around the Hartree-Fock configuration \cite{Robledo2025}, giving the ansatz
\begin{equation}
    \vert\Psi\rangle = e^{-\hat{K}_2}e^{\hat{K}_1}e^{i\hat{J}_1}e^{-\hat{K}_1}\vert\Psi_\text{RHF}\rangle,
\end{equation}
To enforce total-spin conservation in the computed ground states \(\vert\Psi^{(k)}\rangle\), the corrected subspace \(S^{(k)}\) in each batch was decomposed into

\begin{equation}
    U^{(k)} = \bigcup_{\text{x}\in S^{(k)}}\{\text{x}_\alpha, \text{x}_\beta\},
\end{equation}
where x$_\alpha$, x$_\beta$ $\in\{0,1\}^{M/2}$ are the constituent single-spin configurations of length $M/2$ obtained by splitting the corresponding configuration $\text{x}~=~\text{x}_\alpha\text{x}_\beta\in~S^{(k)}$.

The subspace in SQD was then reconstructed as the set of all combinations of unique single-spin configurations,

\begin{equation}
    S^{(k)} = \{\text{ x } \vert \text{ x}=\text{y}\oplus\text{z}, \forall\, \text{y},\text{z} \in U^{(k)}\}.
    \label{eq:tp definition of S}
\end{equation}
As \(U^{(k)}\) does not track the origin of the single-spin configurations \(x_\alpha\) and \(x_\beta\), nor does it distinguish between them, the original \(\alpha\)--\(\beta\) correlations in the sampled configurations generated by the inter-spin Jastrow terms \(\exp(i\hat{J}_{i\alpha\beta})\) in the LUCJ circuit were lost.

Motivated by this we propose the HSQD method, which uses a half-qubit modification of the original LUCJ ansatz as shown in Figure \ref{fig:1}. Due to the spin-symmetric nature of the systems studied in this work, we create a half-qubit circuit for only one spin-type (either \(\alpha\) or \(\beta\), without loss of generality). However, this can be applied to other systems by creating separate half-qubit circuits for each spin-type. The corresponding inter-spin terms $\exp(i\hat{J}_{\mu\alpha\beta})$ are remapped to same-spin terms $\exp(i\hat{J}_{\mu\alpha\alpha})$, halving the number of qubits required for simulation from \(M\) to \(M/2\) and significantly reducing the circuit depth. The resulting sampled subspace of half-configurations then takes the form

\begin{equation}
    \tilde{S} = \{\text{ x } \vert \text{ x}\in\{0,1\}^{M/2}\}.
\end{equation}
By reducing the circuit depth, HSQD mitigates gate errors and qubit decoherence, improving the efficiency of valid configuration sampling, an important achievement for QSCI methods.

\begin{table*}[th]
\centering
\resizebox{\textwidth}{!}{%
\begin{tabular}{lcccccc}
\hline
\textbf{System} & \textbf{Device} & \textbf{Circuit type} & \textbf{\#Qubits} & \textbf{Depth} &
\makecell{\textbf{1-qubit}\\\textbf{gates}} &
\makecell{\textbf{2-qubit}\\\textbf{gates}} \\

\hline
N$_2$ (10e, 26o) & IBM Torino & Full-qubit & 52 & 905 & 10929 & 2023 \\
N$_2$ (10e, 26o) & IBM Torino & Half-qubit & 26 & 429 & 5084 & 855 \\
\hline
$[2\text{Fe-}2\text{S}]$ (30e, 20o) & IBM Pittsburgh & Full-qubit & 40 & 593 & 7687 & 1349 \\
$[2\text{Fe-}2\text{S}]$ (30e, 20o) & IBM Pittsburgh & Half-qubit & 20 & 333 & 3598 & 545 \\
\hline
$[4\text{Fe-}4\text{S}]$ (54e, 36o) & IBM Pittsburgh & Full-qubit & 72 & 1572 & 25810 & 4660 \\
$[4\text{Fe-}4\text{S}]$ (54e, 36o) & IBM Pittsburgh & Half-qubit & 36 & 596 & 11804 & 1777 \\
\hline
\end{tabular}
}
\caption{\textbf{Half-qubit circuits are more noise-robust.} Circuit depth and gate counts for the systems investigated are reported after transpilation and optimization for the IBM quantum devices. This shows  the reduction in depth and number of gates afforded by the half-qubit mapping for LUCJ. All LUCJ circuits use a single layer ($L=1$).}
\label{tab:gatecounts}
\end{table*}

To validate our HSQD method, we first simulated the dissociation of the nitrogen molecule, a well-known benchmark of a method's ability to handle static correlation. First, using a noiseless simulator, HSQD reproduced the PES of N$_2$ with a precision comparable to the original SQD \cite{Robledo2025}. Extending the tests to IBM quantum hardware, the accuracy of HSQD was again comparable to the original SQD, but used 40\% fewer quantum measurements and half the number of qubits. These results are shown in Figure \ref{fig:3} and demonstrate the improved efficiency with which valid electron configurations can be sampled when using the shallower half-qubit ansatz.

Despite this significant improvement over SQD, the computed energies remain above the chemical accuracy threshold of \(1.6\) mHa relative to the reference HCI calculations. We suspect that for realistic applications, the SCCR energies form a loose upper bound for the ground state energy that can be obtained from the corrected subspace \(S\) due to the stochastic nature of the subspace construction in the Monte-Carlo estimator \cite{Robledo2025}. To tighten it, we developed a deterministic HCI-inspired selection procedure that constructs a more optimal selection \(S'\) of the corrected samples $S$, where \(|S'|\ll|S\otimes S|\). We refer to the combination of this HCI-inspired subspace selection and HSQD as HCI-HSQD, and lay out the full workflow in Figure \ref{fig:2}.

Our selection begins from the Hartree–Fock configuration x$_{\text{RHF}}$ and iteratively expands the subspace $S'$ to include new singly and doubly excited configurations x$_j$, constructed from the samples as
\begin{equation}
    \{\text{x}_j = \text{y}\oplus\text{z} \,\vert\, \text{exc}(\text{y},\text{x}_{i\alpha}) + \text{exc}(\text{z},\text{x}_{i\beta}) \in \{1,2\},\text{ y},\text{z}\in S\},
\end{equation}
where x$_i \in S'$ is the state being expanded, exc$(\text{x}_i,\text{ x}_j)$ is the degree of excitation from configuration x$_i$ to x$_j$, and x$_{i\alpha}$ and x$_{i\beta}$ are the corresponding single-spin configurations for x$_i$.
Only configurations x$_j$ satisfying the following condition are selected for $S'$,
\begin{equation}
    \exists\,\text{x}_i\in S': H_{ij}c_i \ge \epsilon_1,
\end{equation}
where $\epsilon_1$ is the HCI variational threshold.
This process continues until either \(S'\) reaches a target size, the energy \(E^{S'}\) has converged, or too few states meet the criteria in each iteration, as illustrated in Figure \ref{fig:2}.

This deterministic selection of the samples accounts for correlations among configurations within \(S\), yielding far more compact and accurate subspaces than the stochastic sampling in SQD. Additionally, by dynamically combining sampled half-configurations to reconstruct full-configurations of length \(M\), we work in the full-qubit space while avoiding the explicit construction of the vast tensor-product subspace $S_\text{tensor}=S\otimes S$.

Figure \ref{fig:4} shows that applying this selection to the post-SCCR subspace \(S\) from HSQD reduces energy errors w.r.t HCI to less than \(1\) mHa across the PES. The resulting subspaces are also 18–39\% smaller than the classically generated subspaces from HCI. This shows that in QSCI methods, whenever the size of the pool of quantum samples necessitates some form of subsampling, improved selection methodologies can produce better energies. It also highlights the value of hybrid quantum-classical frameworks for producing more compact solutions than classical alternatives.

Next, we investigated the dependency of HCI-HSQD on the method used to correct the noisy quantum samples $\tilde{S}$. As SCCR involves numerous diagonalizations of large subspace Hamiltonians \(\hat{H}_{S^{(k)}}\), we evaluated the performance of HCI-HSQD using reference occupations from different intermediate SCCR cycles. We found that correcting the noisy sampled subspace \(\tilde{S}\) using occupations \(\mathbf{n}_v\) obtained from the diagonalization of $\hat{H}$ projected into the valid-sample subspace \(S_v\) is sufficient to maintain energy errors below 1 mHa across the PES with substantially reduced computational cost. Using occupations from later SCCR iterations does provide marginal accuracy improvements but at significantly higher computational expense. These results are shown in Figure \ref{fig:5}.

Lastly, we applied HCI-HSQD to the large-scale simulations of the synthetic iron-sulfur clusters $[2\text{Fe}$–$2\text{S}]$ and $[4\text{Fe}$–$4\text{S}]$ \cite{Robledo2025}. In both cases, we apply a simple probability-based correction to the noisy samples $\tilde{S}$ from the device in place of SCCR, eliminating the need for any correction-related diagonalizations. To highlight the contribution of the shallower half-qubit circuits, we also applied our correction and selection methodologies to the decomposed noisy samples reported in the original SQD work \cite{Robledo2025}, denoting this HCI-SQD. These results are shown in Figure \ref{fig:fes}. 

For $[2\text{Fe}$–$2\text{S}]$, HCI-HSQD and HCI-SQD give identical results, reducing the energy error of SQD relative to the classical DMRG reference by 76\% with a 58\% smaller subspace. However, HCI-HSQD achieves this result with half the number of qubits and quantum measurements as HCI-SQD, highlighting the efficiency gains from using shallower half-qubit circuits. 

For $[4\text{Fe}$–$4\text{S}]$, HCI-SQD again showed an improvement over SQD in accuracy and subspace size as a result of the improved sample selection. However, HCI-HSQD performed even better and reduced the energy error of SQD relative to the classical DRMG reference by 26\% with a 88\% smaller subspace. HCI-HSQD also outperformed the classical HCI method, reducing its energy error relative to the classical DMRG reference by around $1.7$\% with a 76\% smaller subspace. As before, this was achieved using half the number of qubits and quantum measurements as SQD \cite{Robledo2025}.

\section{Discussion}

Achieving quantum advantage with QSCI methods \cite{Kanno2023, Nakagawa2024, Robledo2025, PellowJarman2025, Sugisaki2025, Mikkelsen2025, Yu2025, Shirakawa2025, Yoo2026} depends critically on the ability to efficiently sample the support of the ground-state's electronic wave function. On near-term quantum hardware, this is hindered by some fundamental challenges, including qubit decoherence and gate noise, which distort the prepared quantum state, and wave function broadening from electron correlation, which requires large configuration spaces to accurately capture the wave function support. Configuration recovery methods such as SCCR \cite{Robledo2025} can partially mitigate noise by recovering invalidated samples through symmetry exploitation, yet they cannot fully compensate for inefficient sampling at the hardware level and remain computationally demanding.

In this work, we introduced HSQD, a half-qubit framework for QSCI algorithms that directly targets the sampling inefficiency at its source. By halving the number of qubits required to represent the system, the approach drastically reduces circuit depth. This reduces the accumulation of gate errors and mitigates qubit decoherence, thereby increasing the number of valid configurations in the sampled subspace. We also developed HCI-HSQD, which adds a deterministic selection strategy that tightens the upper bound for the ground-state energy formed by the subspace diagonalizations. Unlike stochastic subspace expansion or basic tensor-product construction, our selection procedure builds compact configuration spaces that retain strong physical correlations.

We demonstrated that simulations of large-scale, strongly correlated systems on NISQ device can be greatly improved by using HCI-HSQD, producing substantially better energies with markedly reduced computational cost. This highlights how improvements in both quantum sampling and classical subspace processing can cooperatively enhance overall performance.

Our application of HCI-HSQD to N$_2$, $[2\text{Fe}$–$2\text{S}]$, and $[4\text{Fe}$–$4\text{S}]$ represents one of the most accurate and resource-efficient QSCI simulations on NISQ hardware reported to date. More broadly, this framework demonstrates a viable route toward practical quantum advantage in QSCI workflows by maximizing the quality of quantum-generated samples and minimizing the need for correction-related post-processing. The reduction in qubit count and circuit depth also extends the applicability of QSCI methods to larger basis sets and more strongly correlated systems within the constraints of NISQ hardware. These results establish a robust foundation for scalable, noise-resilient hybrid simulations of molecular systems and underscore the potential of quantum-centric computation for achieving chemically accurate electronic structure simulations on near-term devices.

\section{Methods}

\begin{figure*}[ht]
    \centering
    \includegraphics[width=0.8\textwidth]{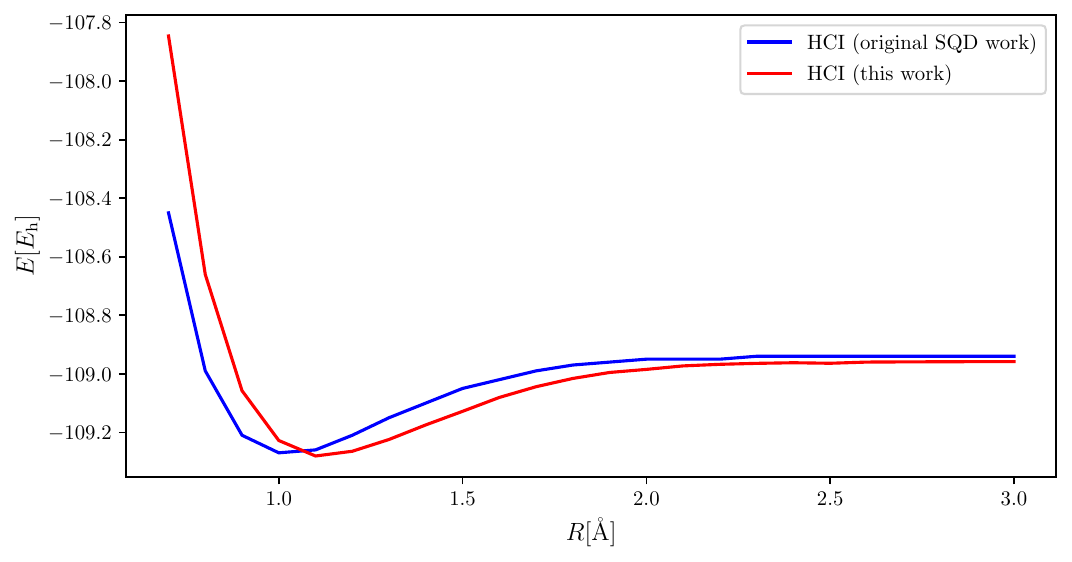}
    \caption{\textbf{Classical references for N$_2$ potential energy surface.} The classical HCI reference for the dissociation of the nitrogen (10e, 26o) system in this work (red) and the original SQD work (blue) \cite{Robledo2025}. As a result of the discrepancy in energies, we were unable to perform a direct comparison of HSQD and SQD for this system. Our HCI reference has been verified by comparing the equilibrium position with that of a more accurate DMRG \cite{White1992, White1993} calculation.}
    \label{fig:hci}
\end{figure*}

All quantum and classical computations reported in this work were performed using the resources and settings described below. Classical diagonalizations were carried out on a CPU cluster consisting of two nodes, each equipped with an Intel Xeon Platinum 8581C processor and 732~GB of RAM. The subspace Hamiltonian diagonalizations and HCI calculations were performed using the DICE SHCI package~\cite{Holmes2016, Smith2017, Sharma2017}. All HCI calculations excluded any sort of perturbative correction, relying only on the variational subspace expansion procedure.

All quantum hardware experiments for the nitrogen system were executed on the IBM Torino device through the IBM Quantum Platform, while experiments for the iron-sulfur clusters were performed on IBM Pittsburgh. Reset mitigation was employed to exclude results arising from incorrect state initialization, and both readout error mitigation and dynamical decoupling were incorporated via the Qiskit Sampler primitive to reduce measurement and decoherence errors.

For each geometry of N$_2$ simulated with HSQD, 300\,000 measurements were performed on the quantum device, while 1\,500\,000 measurements were performed for each iron-sulfur cluster. The SCCR procedure within HSQD for N$_2$ was performed for $10$ cycles, each consisting of $10$ batches and a maximum of 16~million configurations per batch. Upon closer inspection, we found that we were unable to perform a direct comparison of HSQD against SQD for our N$_2$ calculations as the classical calculations reported in the original work \cite{Robledo2025} do not match our own. This is shown in Figure \ref{fig:hci}. We verified our own classical calculations by checking the equilibrium point against reference DMRG calculations.

When sampling from half-qubit circuits in HSQD, the large number of half-states that are produced necessitates some modification to the batch subsampling procedure in original SQD work ~\cite{Robledo2025}. To retain comparable batch parameters, we introduced the concept of a pool $P$, defined as a subset of the noisy samples~$\tilde{S}$ from which the batches in each cycle are subsampled. In all N$_2$ experiments, the valid-sample subspace~$S_v$ was smaller than the defined pool size $\vert P\vert$, so the entirety of~$S_v$ was included in~$P$, and the remaining configurations were drawn from the invalid set~$S_{\mathrm{invalid}}$ according to the probability distribution derived from device measurement counts. This construction provides an upper bound for the ideal subspace and its shortcomings motivated the development of the improved HCI-inspired selection method.

For the original and half-qubit LUCJ circuits transpiled for the IBM devices, we recorded the circuit depth, single-qubit gate count, and two-qubit gate count to quantify the relative efficiency of the two circuit constructions, as shown in Table \ref{tab:gatecounts}.


\begin{thebibliography}{99}

\bibitem{Kanno2023}
Kanno, K., Kohda, M., Imai, R., Koh, S., Mitarai, K., Mizukami, W., Nakagawa, Y. O.
\textit{Quantum-Selected Configuration Interaction: Classical Diagonalization of Hamiltonians in Subspaces Selected by Quantum Computers.}
\textit{arXiv:2302.11320} (2023). \url{https://arxiv.org/abs/2302.11320}

\bibitem{Nakagawa2024}
Nakagawa, Y. O., Kamoshita, M., Mizukami, W., Sudo, S., Ohnishi, Y.-y.
\textit{ADAPT-QSCI: Adaptive Construction of an Input State for Quantum-Selected Configuration Interaction.}
\textit{J. Chem. Theory Comput.} \textbf{20}, 10817–10825 (2024). \url{https://doi.org/10.1021/acs.jctc.4c00846}

\bibitem{Robledo2025}
Robledo-Moreno, J. \textit{et al.}
\textit{Chemistry Beyond the Scale of Exact Diagonalization on a Quantum-Centric Supercomputer.}
\textit{Sci. Adv.} \textbf{11}, eadu9991 (2025). \url{https://doi.org/10.1126/sciadv.adu9991}

\bibitem{PellowJarman2025}
Pellow-Jarman, A. \textit{et al.}
\textit{HIVQE: Handover Iterative Variational Quantum Eigensolver for Efficient Quantum Chemistry Calculations.}
\textit{arXiv:2503.06292} (2025). \url{https://arxiv.org/abs/2503.06292}

\bibitem{Sugisaki2025}
Sugisaki, K. \textit{et al.}
\textit{Hamiltonian Simulation-Based Quantum-Selected Configuration Interaction for Large-Scale Electronic Structure Calculations with a Quantum Computer.}
\textit{Phys. Chem. Chem. Phys.} \textbf{27}, 20869–20884 (2025). \url{https://doi.org/10.1039/D5CP02202A}

\bibitem{Mikkelsen2025}
Mikkelsen, M., Nakagawa, Y. O.
\textit{Quantum-Selected Configuration Interaction with Time-Evolved State.}
\textit{arXiv:2412.13839} (2025). \url{https://arxiv.org/abs/2412.13839}

\bibitem{Yu2025}
Yu, J. \textit{et al.}
\textit{Quantum-Centric Algorithm for Sample-Based Krylov Diagonalization.}
\textit{arXiv:2501.09702} (2025). \url{https://arxiv.org/abs/2501.09702}

\bibitem{Shirakawa2025}
Shirakawa, T., Robledo-Moreno, J., Itoko, T., Tripathi, V., Ueda, K., Kawashima, Y., Broers, L., Kirby, W., Pathak, H., Paik, H., Tsuji, M., Kodama, Y., Sato, M., Evangelinos, C., Seelam, S., Walkup, R., Yunoki, S., Motta, M., Jurcevic, P., Horii, H., Mezzacapo, A.
\textit{Closed-loop Calculations of Electronic Structure on a Quantum Processor and a Classical Supercomputer at Full Scale.}
\textit{arXiv:2511.00224} (2025). \url{https://arxiv.org/abs/2511.00224}

\bibitem{Yoo2026}
Yoo, P., Kim, K., Elala, E. E., McFarthing, S., Pellow, A., Fuks, J. I., Kang, D. H., Nakliang, P., Kim, J., Pathak, H., Shirakawa, T., Yunoki, S., Rhee, J.-K. K.
\textit{Extending the Handover-Iterative VQE to Challenging Strongly Correlated Systems: \texorpdfstring{$\mathrm{N_2}$}{N2} and Fe--S Cluster.}
\textit{arXiv:2601.06935} (2026). \url{https://arxiv.org/abs/2601.06935}

\bibitem{Peruzzo2014}
Peruzzo, A., McClean, J., Shadbolt, P., Yung, M.-H., Zhou, X.-Q., Love, P. J., Aspuru-Guzik, A., O’Brien, J. L.
\textit{A Variational Eigenvalue Solver on a Photonic Quantum Processor.}
\textit{Nat. Commun.} \textbf{5}, 4213 (2014). \url{https://doi.org/10.1038/ncomms5213}

\bibitem{Bartlett1989}
Bartlett, R. J., Kucharski, S. A., Noga, J.
\textit{Alternative Coupled-Cluster Ansätze II. The Unitary Coupled-Cluster Method.}
\textit{Chem. Phys. Lett.} \textbf{155}, 133–140 (1989). \url{https://doi.org/10.1016/S0009-2614(89)87372-5}

\bibitem{Lee2019}
Lee, J., Huggins, W. J., Head-Gordon, M., Whaley, K. B.
\textit{Generalized Unitary Coupled Cluster Wave Functions for Quantum Computation.}
\textit{J. Chem. Theory Comput.} \textbf{15}, 311–324 (2019). \url{https://doi.org/10.1021/acs.jctc.8b01004}

\bibitem{Reinholdt2025}
Reinholdt, P. \textit{et al.}
\textit{Critical Limitations in Quantum-Selected Configuration Interaction Methods.}
\textit{J. Chem. Theory Comput.} \textbf{21}, 6811–6822 (2025). \url{https://doi.org/10.1021/acs.jctc.5c00375}

\bibitem{Holmes2016}
Holmes, A. A., Tubman, N. M., Umrigar, C. J.
\textit{Heat-Bath Configuration Interaction: An Efficient Selected Configuration Interaction Algorithm Inspired by Heat-Bath Sampling.}
\textit{J. Chem. Theory Comput.} \textbf{12}, 3674–3680 (2016). \url{https://doi.org/10.1021/acs.jctc.6b00407}

\bibitem{Smith2017}
Smith, J. E. T., Mussard, B., Holmes, A. A., Sharma, S.
\textit{Cheap and Near-Exact CASSCF with Large Active Spaces.}
\textit{J. Chem. Theory Comput.} \textbf{13}, 5468–5478 (2017). \url{https://doi.org/10.1021/acs.jctc.7b00900}

\bibitem{Sharma2017}
Sharma, S., Holmes, A. A., Jeanmairet, G., Alavi, A., Umrigar, C. J.
\textit{Semistochastic Heat-Bath Configuration Interaction Method: Selected Configuration Interaction with Semistochastic Perturbation Theory.}
\textit{J. Chem. Theory Comput.} \textbf{13}, 1595–1604 (2017). \url{https://doi.org/10.1021/acs.jctc.6b01028}

\bibitem{Xu2025}
Xu, E., Dawson, W., Pathak, H., Nakajima, T.
\textit{Tensor-Product Bitstring Selected Configuration Interaction.}
\textit{arXiv:2503.10335} (2025). \url{https://arxiv.org/abs/2503.10335}

\bibitem{Motta2023}
Motta, M., Sung, K. J., Whaley, K. B., Head-Gordon, M., Shee, J.
\textit{Bridging Physical Intuition and Hardware Efficiency for Correlated Electronic States: The Local Unitary Cluster Jastrow Ansatz for Electronic Structure.}
\textit{Chem. Sci.} \textbf{14}, 11213–11227 (2023). \url{https://doi.org/10.1039/D3SC02516K}

\bibitem{Sharma2014}
Sharma, S., Sivalingam, K., Neese, F., Chan, G. K.-L.
\textit{Low-energy spectrum of iron–sulfur clusters directly from many-particle quantum mechanics.}
\textit{Nat. Chem.} \textbf{6}, 927–933 (2014). \url{https://doi.org/10.1038/nchem.2041}

\bibitem{White1992}
White, S. R.
\textit{Density matrix formulation for quantum renormalization groups.}
\textit{Phys. Rev. Lett.} \textbf{69}, 2863–2866 (1992). \url{https://link.aps.org/doi/10.1103/PhysRevLett.69.2863}

\bibitem{White1993}
White, S. R.
\textit{Density-matrix algorithms for quantum renormalization groups.}
\textit{Phys. Rev. B} \textbf{48}, 10345–10356 (1993). \url{https://link.aps.org/doi/10.1103/PhysRevB.48.10345}


\end{thebibliography}
\end{document}